\documentclass[aip,amsmath,amssymb,reprint]{revtex4-1}

\usepackage{graphicx}

\begin{document}

\title{Terahertz active spatial filtering through optically tunable hyperbolic metamaterials}

\author{Carlo Rizza}
\affiliation{Dipartimento di Scienza e Alta Tecnologia, Universit\`a dell'Insubria, Via Valleggio 11, 22100 Como, Italy} \affiliation{Consiglio Nazionale
delle Ricerche, CNR-SPIN, 67100 Coppito L'Aquila, Italy}

\author{Alessandro Ciattoni}
\email{alessandro.ciattoni@aquila.infn.it}
\affiliation{Consiglio Nazionale delle Ricerche, CNR-SPIN, 67100 Coppito L'Aquila, Italy}

\author{Elisa Spinozzi}
\address{University of Rome ``La Sapienza'', Department of Information Engineering Electronics and Telecommunications, Via Eudossiana 18, 00184 Roma,
Italy}

\author{Lorenzo Columbo}
\affiliation{Dipartimento di Scienza e Alta Tecnologia, Universit\`a  dell'Insubria, Via Valleggio 11, 22100 Como, Italy} \affiliation{Consiglio Nazionale
delle Ricerche, CNR-IFN, 70126 Bari, Italy}

\begin{abstract}
We theoretically consider infrared-driven hyperbolic metamaterials able to spatially filtering terahertz radiation. The metamaterial is a slab made of
alternating semiconductor and dielectric layers whose homogenized uniaxial response, at terahertz frequencies, shows principal permittivities of different
signs. The gap provided by metamaterial hyperbolic dispersion allows the slab to stop spatial frequencies within a bandwidth tunable by changing the
infrared radiation intensity. We numerically prove the device functionality by resorting to full wave simulation coupled to the dynamics of charge carries
photoexcited by infrared radiation in semiconductor layers.
\end{abstract}

\maketitle

Manipulating terahertz (THz) radiation is generally a difficult task since the most of standard materials simply do not respond to such frequencies.
However, the advent of metamaterials has allowed to partially reduce this difficulty since their electromagnetic properties can be artificially
manipulated \cite{Meta2} through a suitable design of the underlying constituent unit cells. At the same time a number of setups have been proposed for
steering the THz radiation \cite{Setu2} and reconfigurable electrically \cite{MeEl2} or optically \cite{MeOp1,MeOp3} driven metamaterials have been
exploited for conceiving active THz devices \cite{Devi1}.

The most of the proposed active THz devices are tunable frequency-domain filters since it is relatively simple to control the metamaterial dispersion
properties through external stimuli. In this Letter we theoretically propose a way for achieving {\it active} and {\it spatial} filtering of the THz
radiation by means of a suitable hyperbolic metamaterial whose THz response can be tuned by an auxiliary infrared field. Hyperbolic or indefinite media
\cite{Smit1} are uniaxially anisotropic metamaterials having principal permittivities of different signs, a remarkable feature leading extraordinary plane
waves to be ruled by a hyperbolic dispersion relation. Hyperbolicity is the main physical ingredient leading to unusual optical effects as negative
refraction \cite{Smit2} and hyperlensing \cite{Jacob} and supporting a number of proposed devices as beam splitters \cite{Zhao1}, spatial \cite{Shuri} and
angular filters \cite{Aleks} and optical switches \cite{Ciat1}. The tunable hyperbolic metamaterial we consider in the present Letter, together with the
fields geometry, is sketched in Fig.1. The metamaterial slab of thickness $L$ is obtained by stacking along the $x$-axis alternating layers of an
intrinsic semiconductor \cite{Hoffm} (sc) and a negative dielectric (nd) of thicknesses $d_{sc}$ and $d_{nd}$, respectively and it is illuminated by an
infrared (IR) plane wave linearly polarized along the $y$-axis and normally impinging onto the slab interface at $z=0$. The THz field (TH) is a transverse
magnetic (TM or p-polarized) monochromatic plane impinging with incidence angle $\theta$ onto the interface.

\begin{figure}
\includegraphics*[width=0.45\textwidth]{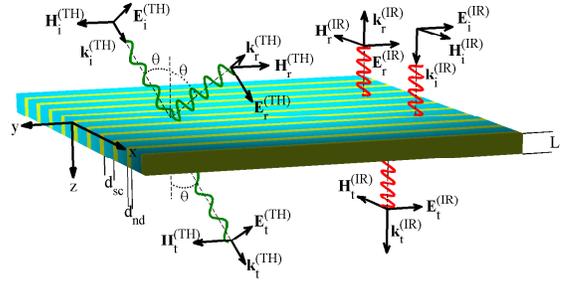}
\caption{(Color online) Layered metamaterial setup together with THz (TH) and infrared (IR) plane waves geometry.}
\end{figure}

The infrared field within the semiconductor layers photoexcites electrons to the conduction band which dynamically recombine so that the resulting
electron density $N$ is described by the rate equation \cite{Garmi}
\begin{equation} \label{carriers}
\frac{\partial N}{\partial t} = \frac{\epsilon_0}{2 \hbar}{\rm Im}\left[\epsilon_{sc}(\omega_{IR})\right] |E^{(IR)}|^2-\frac{N}{\tau_R}-BN^2,
\end{equation}
where $\hbar$ is the Planck constant divided by $2\pi$, $\epsilon_0$ is the absolute vacuum permittivity, $\epsilon_{sc}(\omega_{IR})$ is the
semiconductor permittivity at the infrared frequency $\omega_{IR}$, $E^{(IR)}$ is the infrared field within the semiconductor layers, $\tau_R$ is the
carriers' nonradiative recombination time and B is the coefficient of radiative recombination.  The semiconductor dielectric behavior at the infrared
frequency is described by the linearized permittivity model \cite{Garmi}
\begin{equation} \label{epsilonIR}
\epsilon_{sc}(\omega_{IR}) = \left( n_{IR} + \frac{\delta n_0}{N_0} N  \right)^2 + i \frac{n_{IR} c }{\omega_{IR}} A(N_0-N)
\end{equation}
where $n_{IR}$ is the infrared semiconductor refractive index background, $N_0$ is the transparency value of the carrier density, $\delta n_0$ the
refractive index change at transparency, $c$ is the speed of light in vacuum and A is the differential absorption coefficient. At steady state ($\partial
N/\partial t = 0$), Eqs.(\ref{carriers}) and (\ref{epsilonIR}) yield
\begin{eqnarray} \label{density}
N = \frac{1}{2\tau_R B} \left[ -\left( 1 + W \right) + \sqrt{ \left( 1 + W \right) ^2 + 4 \tau_R B N_0 W} \right]
\end{eqnarray}
where $W = |E^{(IR)}|^2 / |E_{sat}|^2$ and $|E_{sat}| = \left[ \tau_R \epsilon_0 n_{IR} c A / (2 \hbar \omega_{IR}) \right]^{-1/2}$, so that electron
density shows a saturable behavior (i.e. $N \simeq N_0$ if $|E^{(IR)}| \gg |E_{sat}|$), as a consequence of the saturable absorption model of
Eq.(\ref{epsilonIR}). Maxwell equations for the field $E^{(IR)}$ together with Eqs.(\ref{epsilonIR}) and (\ref{density}) describe the infrared nonlinear
behavior within semiconductor layers. The resulting (and self-consistently evaluated) electron density $N$ has a strong impact on the semiconductor
response at the THz frequency $\omega_{TH}$, since the permittivity is
\begin{equation} \label{Drude}
\epsilon_{sc}(\omega_{TH}) = \epsilon_{sc}^{(0)}(\omega_{TH})+\frac{i}{\epsilon_0} \frac{(e^2 \tau /m^*) N}{\omega_{TH} (1-i\omega_{TH}\tau)}
\end{equation}
where $\epsilon_{sc}^{(0)}(\omega_{TH})$ is the semiconductor permittivity in the absence of the infrared radiation, $-e$ and $m^*$ are electron charge
and reduced mass and $\tau$ is the relaxation time. Note that the term proportional to $N$ in Eq.(\ref{Drude}) is the standard Drude permittivity
contribution due to conduction band electrons and it is the physical ingredient allowing the infrared field to tune the overall slab THz response. If the
layers thicknesses $d_{sc}$ and $d_{nd}$ are much smaller than the THz wavelength (of the order of tens of microns), the overall metamaterial slab of
Fig.1 shows a homogeneous THz uniaxial response with permittivities (the optical axis lying along the staking $x$-axis) \cite{Elser}
\begin{eqnarray} \label{homog}
\epsilon_x(\omega_{TH}) &=& \left[ \frac{f_{sc}}{\epsilon_{sc}(\omega_{TH})} + \frac{f_{nd}}{\epsilon_{nd}(\omega_{TH})} \right]^{-1}, \nonumber \\
\epsilon_z(\omega_{TH}) &=& \epsilon_y (\omega_{TH}) = f_{sc} \epsilon_{sc}(\omega_{TH}) + f_{nd} \epsilon_{nd}(\omega_{TH})
\end{eqnarray}
\begin{figure}
\includegraphics*[width=0.45\textwidth]{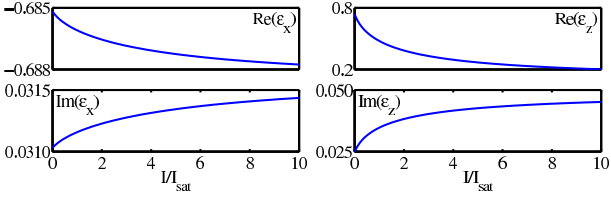}
\caption{(Color online) Terahertz homogenized dielectric permittivities of Eqs.(\ref{homog}) as functions of the normalized local optical intensity of the
infrared field.}
\end{figure}
where $f_{sc} = d_{sc}/(d_{sc}+d_{nd})$ and $f_{nd} = d_{nd}/(d_{sc}+d_{nd})$ are the layers filling fractions whereas $\epsilon_{nd}(\omega_{TH})$ is the
the dielectric permittivity of the negative dielectric layers for which ${\rm Re} \left[\epsilon_{nd}(\omega_{TH})\right]<0$.

It is well known that, since the permittivities of Eqs.(\ref{homog}) involve averages of different kinds, it is possible to tailor the structure in such a
way that $\epsilon_x$ and $\epsilon_z$ have different signs, the resulting hyperbolic medium begin tunable, in the present analysis, through the auxiliary
infrared field. In Fig.2 we have plotted $\epsilon_x$ and $\epsilon_z$ evaluated from Eqs.(\ref{homog}) as functions of the infrared optical intensity
$I^{(IR)} = \epsilon_0 c |E^{(IR)}|^2 /2$ (normalized with the saturation intensity $I_{sat}= \epsilon_0 c |E_{sat}|^2 /2 = 3.7 \: kW/cm^2$), with the
simplified assumption that such intensity can be regarded as uniform within the bulk of the slab and equal to incident infrared plane wave intensity (see
below). The considered wavelengths are $\lambda_{IR} = 2\pi c /\omega_{IR} = 0.879 \: \mu m$ and $\lambda_{TH} = 2\pi c /\omega_{TH} = 23.08 \: \mu m$ and
we have chosen gallium arsenide (GaAs) as semiconductor (for which $\tau_R = 10 \: ns$, $B=1.3 \cdot 10^{-10} \: cm^3 s^{-1}$, $n_{IR} = 3.6$, $\delta n_0
= -0.07$, $N_0 = 9.91 \cdot 10^{17} \: cm^{-3}$, $A=1.71 \cdot 10^{-15} \: cm^{2}$, $\epsilon_{sc}^{(0)}(\omega_{TH}) = 12.9$, $\tau = 3.29 \cdot 10^{-13}
s$ and $m^* = 0.067 m_0$, $m_0$ being the electron mass) with filling fraction $f_{sc}=0.1$ and a negative dielectric with permittivity
$\epsilon_{nd}(\omega_{TH}) = -0.61+0.003i$ (this value coinciding with that of calcium fluoride $\rm CaF_2$ with imaginary part reduced by a tenth for
discussion purposes). It is evident from Fig.2 that the structure has been tailored to show indefinite permittivity tensor in the absence of infrared
illumination, i.e. ${\rm Re}(\epsilon_x)<0$ and ${\rm Re}(\epsilon_z)>0$ for $I^{(IR)}=0$.
\begin{figure}
\includegraphics*[width=0.45\textwidth]{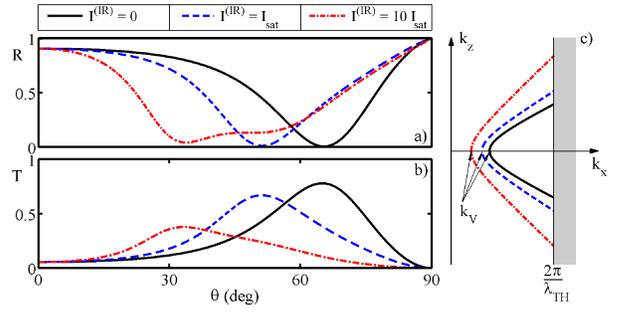}
\caption{(Color online) a) Reflectivity R and b) transmissivity T at the wavelength $\lambda_{TH} = 23.08 \: \mu m$ as functions of the THz incidence
angle $\theta$ for a homogenized slab of thickness $L=9.23 \: \mu m$ and permittivities of Fig.1 at three different {\it local} infrared intensities
$I^{(IR)}$. c) Terahertz hyperbolic dispersion curves for the three cases of panels a) and b). The hyperbola vertex is at $k_x = k_V = (2\pi
/\lambda_{TH})\sqrt{|\epsilon_z|}$}
\end{figure}
Note that the homogenized structure keeps its indefinite character for increasing $I^{(IR)}$ and the permittivities globally show saturation at high
intensities as a consequence of the carrier density saturation predicted by Eq.(\ref{density}). Remarkably, the infrared intensity $I^{(IR)}$ has a strong
impact on ${\rm Re}(\epsilon_z)$. In Fig.3 we have plotted the reflectivity $R=\left|{\bf E}_r^{(TH)}\right|^2/\left|{\bf E}_i^{(TH)}\right|^2$ and the
transmissivity $T=\left|{\bf E}_t^{(TH)}\right|^2/\left|{\bf E}_i^{(TH)}\right|^2$ (see Fig.1 for the definition of the field amplitudes) as functions of
the THz incidence angle $\theta$ for a homogenized slab of thickness $L=9.23 \: \mu m$ with permittivities coinciding with those of Fig.2 at the
intensities $I^{(IR)}=0$, $I^{(IR)} = I_{sat}$ and $I^{(IR)} = 10 I_{sat}$. It is worth noting that all the three transmissivity curves have a bell-shaped
profile with maximum located at an angle dependent on the infrared intensity (analogously the reflectivity have a complementary behavior) so that the slab
can be regarded as an active spatial filter allowing (forbidding) transmission (reflection) of those THz plane waves with incidence angle close to a
central angle in turn tunable through the infrared optical intensity.

The physical mechanism supporting such an active spatial filtering functionality can easily  be grasped by noting that the chosen TM incident THz plane
waves couples to the slab extraordinary waves whose dispersion relation is $k_x^2/|\epsilon_z| - k_z^2/|\epsilon_x| = (2 \pi / \lambda_{TH})^2$, where the
signs of the permittivities (from Fig.2) have been explicitly reported for clarity purposes (neglecting permittivities' imaginary parts). Such dispersion
relation is, in the $k_xk_z$ plane, a hyperbola with vertices along the $k_x$-axis located at $k_x = \pm k_V$ where $k_V = (2 \pi / \lambda_{TH})
\sqrt{|\epsilon_z|} $ (see panel (c) of Fig.3 where the hyperbola corresponding to the considered three infrared intensities are plotted). Momentum match
at the interface $z=0$ implies that the extraordinary plane waves have transverse wave vector $k_x = (2 \pi / \lambda_{TH}) \sin \theta$ so that in the
two situations $|\sin \theta| < \sqrt{|\epsilon_z|}$  or $|\sin \theta| > \sqrt{|\epsilon_z|}$ (i.e. $|k_x|<k_V$ or $|k_x|>k_V$, respectively) the
externally impinging THz plane wave excites, within the slab, evanescent or propagating waves, respectively. As a consequence, the transmission is low for
$\sin \theta < \sqrt{|\epsilon_z|}$ (where evanescent waves provides a residual "tunnelling" radiation), it reaches a maximum at greater angles (due to
the slab propagating waves) and it eventually vanishes at $\theta = 90^\circ$ for geometrical reasons.
\begin{figure}
\includegraphics*[width=0.45\textwidth]{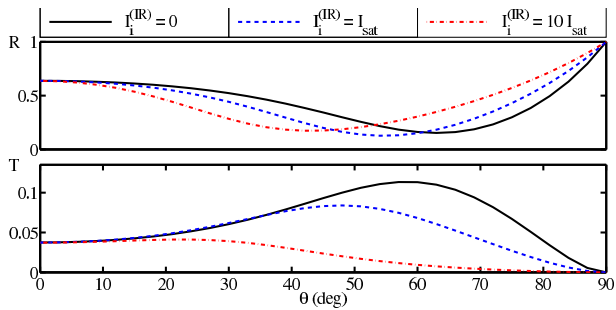}
\caption{(Color online) THz transmissivity R and reflectivity T of the metamaterial of Fig.1, evaluated trough full waves simulations, for three different
values of the optical intensity $I^{(IR)}_i$ of the {\it incident} infrared plane wave.}
\end{figure}

The discussion of the THz spatial filtering functionality has hitherto been based, for clarity purposes, on three simplified assumption i.e 1) low loss
regime (small imaginary part of the negative dielectric permittivity), 2) THz electromagnetic homogenization of the layered medium (validity of
Eqs.(\ref{homog})) and 3) uniformity of the infrared optical intensity within the slab bulk. While the first two assumptions can easily be supported, the
third one is more serious since the layers periodicity is generally comparable with the infrared wavelength and, in addition, reflection of infrared
radiation by the slab together with its nonlinear behavior within semiconductor layers have to be considered. In order to account for of all these
physical ingredients and to show that the structure actually behaves as an active THz spatial filter, we have resorted to full wave simulations where
linear and nonlinear Maxwell equations for the THz and the infrared field are coupled to the electron dynamics described by Eqs.(\ref{density}) and
(\ref{Drude}). We have chosen the layers thicknesses $d_{sc}=24 \: nm$ and $d_{nd}=220 \: nm$ and the permittivities $\epsilon_{nd}(\omega_{TH}) =
-0.61+0.03i$, $\epsilon_{nd}(\omega_{IR}) = 2.04$ which are the $\rm CaF_2$ permittivities at the considered THz and infrared frequencies, respectively,
whereas all the remaining involved parameters are those used above. The results of the simulations are reported in Fig.4 where we plotted the THz
transmissivity and reflectivity of the structure for three different values of the optical intensity $I^{(IR)}_i = \epsilon_0 c |{\bf E}^{(IR)}_i|^2 /2$
of the {\it incident} infrared plane wave (see Fig.1). Note that the overall spatial filtering functionality is evidently exhibited by the considered
realistic structure, even though the values of the transmissivity are smaller as compared to those of Fig.3 as a consequence of the $\rm CaF_2$
absorption.

In conclusion we have proposed a metamaterial structure which, driven by an auxiliary infrared field, is able to spatially filtering THz radiation. We
believe that the proposed combination of hyperbolic metamaterial and semiconductor concepts can suggest different ideas for conceiving novel and efficient
THz active devices.

This research has been funded by the Italian Ministry of Research (MIUR) through the "Futuro in Ricerca" FIRB-grant PHOCOS - RBFR08E7VA. The authors
acknowledge useful discussions with Prof. Massimo Brambilla.

\bibliography{paper}

\end{document}